\documentclass[conference]{IEEEtran}
\IEEEoverridecommandlockouts
\usepackage{cite}
\usepackage{amsmath,amssymb,amsfonts}
\usepackage{algorithmic}
\usepackage{graphicx}
\usepackage{textcomp}
\usepackage{xcolor}
\usepackage{hyperref}
\usepackage[caption=false,font=footnotesize]{subfig}
\usepackage{stfloats}

\usepackage{tikz}
\pgfdeclarelayer{foreground}
\pgfsetlayers{background, main, foreground}

\newcommand\copyrighttext{%
    \footnotesize \begin{center} \color{red} \textcopyright\,2025 IEEE. Personal use of this material is permitted. Permission from IEEE must be obtained for all other uses, in any current or future media, including reprinting/republishing this material for advertising or promotional purposes, creating new collective works, for resale or redistribution to servers or lists, or reuse of any copyrighted component of this work in other works. \end{center}}

\begin{document}

\title{\vspace{-2em} \copyrighttext \huge Doing More With Less: Towards More Data-Efficient Syndrome-Based Neural Decoders\\
\thanks{This work has been funded by the French National Research Agency AI4CODE project (Grant ANR-21-CE25-0006). Part of it was performed using HPC resources from GENCI-IDRIS (Grant 2025-AD011016057)}
}

\author{\IEEEauthorblockN{Ahmad Ismail, Raphaël Le Bidan, Elsa Dupraz, Charbel Abdel Nour}
\IEEEauthorblockA{\textit{IMT Atlantique, Lab-STICC UMR CNRS 6285, Brest, France} \\
\textit{IMT Atlantique, Dept. MEE}\\
\textit{Technopole Brest Iroise, CS 83818, 29238 Brest Cedex 3, France}\\
Email: name.surname@imt-atlantique.fr}
}

\maketitle

\begin{abstract}
While significant research efforts have been directed toward developing more capable neural decoding architectures, comparatively little attention has been paid to the quality of training data. In this study, we address the challenge of constructing effective training datasets to maximize the potential of existing syndrome-based neural decoder architectures. We emphasize the advantages of using fixed datasets over generating training data dynamically and explore the problem of selecting appropriate training targets within this framework. Furthermore, we propose several heuristics for selecting training samples and present experimental evidence demonstrating that, with carefully curated datasets, it is possible to train neural decoders to achieve superior performance while requiring fewer training examples. Code to reproduce all results is available at \href{https://github.com/lebidan/sbnd}{https://github.com/lebidan/sbnd}.
\end{abstract}


\section{Introduction}

The remarkable achievements of deep learning in areas like computer vision and natural language processing have sparked significant interest in applying this approach to complex tasks in other fields.
Channel decoding is no exception \cite{gruber2017}. Various neural decoder architectures have been proposed to address the limitations of traditional decoding algorithms and achieve enhanced performance \cite{matsumine2024}.
In this work, we focus on the challenge of approaching the performance of optimal soft-decision Maximum-Likelihood Decoding (MLD) for short block codes using neural decoders. Our investigation is guided by several key questions: Are current neural architectures capable of achieving near-MLD performance for codes of practical relevance? If so, how does the complexity of the model scale with parameters such as code length or minimum distance? Additionally, how does the performance of these architectures evolve as a function of the size of the training set? 

In the literature, two main categories of neural decoders are typically distinguished. \emph{Model-based} decoders leverage graphical representations and message-passing algorithms originally designed for long sparse-graph codes, to develop decoders that also perform effectively with more general linear codes. Notable contributions in this area include the Neural Belief Propagation (NBP) decoder and its various extensions \cite{nachmani2018,buchberger2021,rosseel2022}, as well as the Graph Neural Network (GNN) decoder \cite{cammerer2022,tian2023}. In contrast, \emph{model-free} decoders encompass general decoding architectures that have minimal dependence on the algebraic or graphical structure of the code. A significant milestone in this category is the syndrome-based neural decoding framework (SBND), initially proposed in \cite{bennatan2018} and further refined in \cite{debonirovella2023,artemasov2023}. A specific instance of SBND is the Error Correction Code Transformer (ECCT), based on the Transformer architecture, introduced in \cite{choukroun2022}, with further enhancements subsequently described in \cite{choukroun2024,park2024}.
Our study focuses on SBND decoders, as these models exhibit less inductive bias compared to NBP decoders, giving them greater potential to approach MLD.

The majority of prior research on neural decoders, including most references cited in this paper, employs on-demand data generation to create training examples dynamically. This approach can be traced back to early work on deep learning-based decoders; see, e.g., \cite{gruber2017}. This practice contrasts with the standard deep learning approach of using fixed reference datasets and complicates the benchmarking of models under consistent and fair conditions. Additionally, it makes it more challenging to estimate the amount of training data needed to reach a specific performance level. Notably, many published models appear to have been trained on exceptionally large datasets, even for relatively small codes. This raises the question of whether such dataset sizes are truly necessary, especially given the scale of some of these models.

In the field of deep learning, it is increasingly acknowledged that not all training samples contribute equally to model learning \cite{Mirzasoleiman2024}. The rise of the Transformer architecture, which has significantly increased both model size and the volume of data required for training, has made the problem even more acute. Consequently, there is a growing emphasis on curating large-scale datasets not only to reduce training costs but also to enhance the generalization capabilities of models. Examples of proposed solutions include learning the most relevant features from compressed datasets \cite{gribonval2021}, subsampling meaningful examples from training data \cite{kolossov2024}, and modifying the distribution of the training set~\cite{nguyen2024}. In the context of neural decoding, \cite{beery2020} was among the first to highlight the importance of high-quality training data and to advocate for smart sampling strategies to improve the efficiency of training model-based NBP decoders.

The central premise of this work is that, to date, insufficient attention has been devoted to the training data in the field of neural decoding. Our primary contribution is to demonstrate that, by using carefully designed fixed datasets, existing neural decoders can be trained more efficiently to achieve comparable or even better performance than before, using fewer examples. 

The paper is organized as follows. Section \ref{sec:prelim} introduces the transmission system and the decoder models. Section \ref{sec:on-demand} describes the on-demand training paradigm. In Section~\ref{sec:datasets}, we highlight the importance of using curated datasets for training and provide design principles and heuristics to construct datasets that achieve superior test performance with fewer samples compared to on-demand training. Experimental results demonstrating the advantages of the proposed training methodology are presented in Section \ref{sec:experiments}. Finally, Section \ref{sec:conclusion} summarizes the key findings of this work.

\section{Preliminaries}
\label{sec:prelim}

\subsection{Transmission model}

\begin{figure}[t!]
\centerline{\includegraphics[scale=0.7]{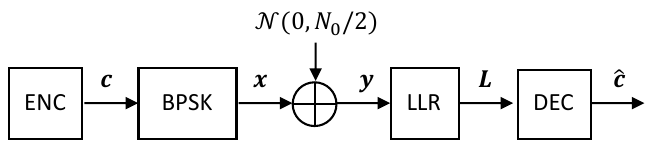}}
\caption{Transmission system model.}
\label{fig:tx_model}
\end{figure}

The transmission model under consideration is depicted in Fig. \ref{fig:tx_model}. Let $\mathcal{C}$ be an $(n, k)$ binary linear code of length $n$ and dimension $k$, and $\mathbf{H}$ be a parity-check matrix for $\mathcal{C}$. A codeword $\mathbf{c}=(c_1, \ldots, c_n)$ is modulated into the Binary-Phase-Shift Keying (BPSK) sequence $\mathbf{x}=(x_1, \ldots, x_n)$, using the mapping $x = (-1)^c$, and transmitted over a Memoryless Binary-Input Output-Symmetric (MBIOS) channel. Let $\mathbf{y}=(y_1, \ldots, y_n)$ be the received word. Assume that the samples $y$ at the output of the channel are real-valued, with conditional probability density functions $f_0(y) = f(y|x=+1)$ and $f_1(y) = f(y|x=-1)$, and denote by $L_i = \log \frac{f_0(y_i)}{f_1(y_i)}$ the Log-Likelihood Ratio (LLR) for code bit $i$. A soft-decision channel decoder aims to infer the transmitted codeword from the LLR vector $\mathbf{L}=(L_1, \ldots, L_n)$. Here, the focus will be placed on the Binary-Input Additive White Gaussian Noise (BI-AWGN) channel with variance $\sigma^2 = \frac{N_0}{2}$, for which $L_i = \frac{2}{\sigma^2} y_i$. 

\subsection{Optimal decoder}

Assuming equiprobable codewords, it is well known (see, e.g. \cite{snyders1989}) that the optimal MLD rule consists in finding a codeword $\mathbf{c} \in \mathcal{C}$ that maximizes the correlation metric $\langle \mathbf{c}, \mathbf{L} \rangle = \sum_{i=1}^n (-1)^{c_i} L_i$ . Equivalently, one may take a hard-decision $z_i$ on each received sample with
\begin{equation*}
z_i = \begin{cases} 0 & y_i \ge 0 \\ 1 & y_i < 0 \end{cases}\;, 
\end{equation*}
calculate the syndrome $\mathbf{s}=\mathbf{z}\mathbf{H}^t$, and decode to $\hat{\mathbf{c}} = \mathbf{z} - \mathbf{e}$ where $\mathbf{e}$ is the most likely error pattern, \emph{i.e.} the pattern with minimum reliability weight \cite{berlekamp1983,snyders1989}
\begin{equation}
w_L(\mathbf{e}) = \sum_{i:e_i=1} |L_i|
\end{equation}
within the coset indexed by $\mathbf{s}$.
It follows that both the vector \(\mathbf{L}\) and the pair \((\mathbf{s}, |\mathbf{L}|)\) can serve as sufficient statistics for decoding, where \(|\mathbf{L}|\) denotes the bit reliability vector \((|L_1|, \ldots, |L_n|)\). The correlation-based and syndrome-based formulations of the MLD rule described above are applicable to any MBIOS channel. In general, both decoding approaches have \(O(2^k)\) time complexity, unless specific structure in the code \(\mathcal{C}\) can be leveraged to simplify one approach or the other.

\subsection{Syndrome-based deep learning decoders}

\begin{figure}[tbp]
\centerline{\includegraphics[scale=0.7]{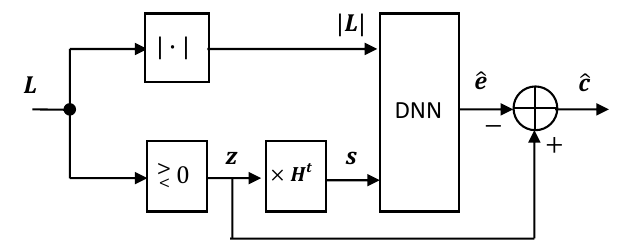}}
\caption{General architecture of a syndrome-based neural decoder.}
\label{fig:sbnd}
\end{figure}

In this work, we focus on training SBND decoders to approximate the MLD rule and achieve near-MLD performance at a reduced computational cost. The general structure of an SBND decoder, as introduced by \cite{bennatan2018}, is depicted in Fig.~\ref{fig:sbnd}. Using the input vector pair $(\mathbf{s}, |\mathbf{L}|)$, the SBND decoder infers the most likely binary error pattern $\hat{\mathbf{e}}$ within the coset corresponding to the hard decision $\mathbf{z}$ on the received word. The core component of the decoder, denoted as DNN, is a Deep Neural Network in charge of estimating $\hat{\mathbf{e}}$. In this study, we consider two distinct DNN architectures: the recurrent network composed of stacked Gated Recurrent Units (GRU) introduced in \cite{bennatan2018} and the Transformer-based ECCT model proposed in \cite{choukroun2022}. These two SBND models have garnered the most attention in the literature to date. However, we stress that the ideas presented here are not restricted to these models and can be applied to other model-free decoders as well.

For the sake of conciseness, and because this paper focuses on training data and methodology, we do not provide detailed descriptions of the two architectures. Instead, we refer interested readers to the original works \cite{bennatan2018} and \cite{choukroun2022}. Here, we introduce the notations used to reference the models throughout the paper. The stacked GRU model has three hyperparameters: the hidden layer size $h$, the number of layers $\ell$, and the number of time steps $t$. Consistent with \cite{bennatan2018}, we fix the hidden size to $h=6(2n-k)$. Accordingly, $\text{GRU}(\ell,t)$ denotes a stacked GRU model with $\ell$ layers, $t$ time steps, and $h=6(2n-k)$ hidden units. Similarly, the ECCT model has three main hyperparameters: the embedding dimension $d$, the number of attention layers $\ell$, and the number of self-attention heads $a$. Following \cite{choukroun2022}, we fix $a=8$ in this work. Therefore, $\text{ECCT}(d,\ell)$ refers to an ECCT model with embedding size $d$, $\ell$ layers, and $a=8$ self-attention heads.

\subsection{SBND models implementation, training, and testing}

We used PyTorch to implement, train, and evaluate the models discussed in this work. This subsection provides detailed information regarding the implementation process. 

\subsubsection{Model input}
Since any scaled version of the LLR vector $\mathbf{L}$ serves as a sufficient statistic for decoding, we use the scaled reliability vector $|\mathbf{y}|$ in place of $|\mathbf{L}|$ as the input to the SBND architecture in Fig.~\ref{fig:sbnd}. Furthermore, to maintain the reliability values $|y_i|$ within the range $[0,1]$, we normalize them with respect to the maximum reliability value in each received word, since this normalization does not alter the MLD decision. No additional normalization is applied to the input data. The syndrome $\mathbf{s}$ is represented in bipolar form rather than as a bit vector, as we observed a slight improvement in training under this representation. This observation aligns with similar conclusions drawn by other authors \cite{choukroun2022, debonirovella2023}.

\subsubsection{Core DNN implementation}
Our stacked GRU decoder is built using PyTorch's GRU model. We found that the GRU bias terms were unnecessary, which allowed us to reduce the number of parameters. For the ECCT model, we used the source code available in \cite{github_ecct}, retaining the default dropout rate of $0.1$ at the output of the multi-head attention layer.

\subsubsection{Model output}
From the pair $(\mathbf{s}, |\mathbf{y}|)$, the SBND model in Fig.~\ref{fig:sbnd} computes a real-valued estimate $\hat{\mathbf{e}}$ of the error pattern. A tanh function is applied at the output of the core DNN to squash the activations $\hat{e}_i$ within the range $[-1, +1]$. These activations can then be turned into bit probability estimates through the inverse BPSK mapping $\Pr(e_i = 1) \approx (1 - \hat{e}_i) / 2$.

\subsubsection{SBND model training}

Syndrome-based MLD can be formulated as a multiclass classification problem, specifically selecting the optimal error pattern $\mathbf{e}$ from the $2^k$ candidate error patterns that comprise the coset $\mathbf{z}+\mathcal{C}$, with the goal of minimizing the reliability weight $w_L(\mathbf{e})$. Thus, minimizing the cross-entropy loss would appear to be the most appropriate criterion for training SBND models in a supervised manner. However, the large number of possible classes makes this approach computationally intractable. To overcome this challenge, the commonly adopted solution is to minimize the Binary Cross-Entropy (BCE) loss instead, which is defined as:
\begin{equation*}
\mathcal{L}(\mathbf{e},\hat{\mathbf{e}})  = \frac{1}{n} \sum_{i=1}^n - e_i \log_2\left(\frac{1-\hat{e_i}}{2}\right)  - (1-e_i) \log_2\left(\frac{1+\hat{e_i}}{2}\right) 
\end{equation*}
This loss is computed between the binary target error pattern $\mathbf{e}$ and the activation vector $\hat{\mathbf{e}}$ at the model's output. By employing this approach, the overwhelming multiclass error pattern classification problem is effectively reduced to a more manageable bit-by-bit binary classification task. The selection of target error patterns $\mathbf{e}$ for training purposes will be discussed in Sections \ref{sec:on-demand} and \ref{sec:datasets}. At this stage, it is important to emphasize that the training data only include received words with a non-zero syndrome, as these are the only words that will be passed to the decoder in a practical setting. Training is conducted at a single, carefully chosen Signal-to-Noise Ratio (SNR) value. Additional details regarding the training setup are provided in the Appendix. 

\subsubsection{SBND model evaluation}
In line with standard channel coding practices, we use the Frame Error Rate (FER) to assess the performance of the model during the testing phase. Specifically, a frame error is recorded when the estimated error pattern $\hat{\mathbf{e}}$ differs from the true error pattern $\mathbf{e} = \mathbf{z} - \mathbf{c}$ by at least one bit at the output of the decoder. The test FER is calculated through Monte Carlo simulations of the model operating in inference mode, across various SNR levels. While only the all-zero codeword is employed during training, the testing phase is conducted on randomly generated codewords to provide a more comprehensive evaluation of model performance.

\section{The Training With On-Demand Data Paradigm}
\label{sec:on-demand}

\subsection{What does training with on-demand data mean?}

Training with on-demand data involves generating new batches of noisy received words at each step of the training, resulting in a dataset that is dynamically created as training proceeds. In this scenario, the concept of an epoch technically loses relevance, as the model never encounters the same training example twice. On-demand data generation aligns naturally with the Monte Carlo simulation approach widely used in research and design for communication systems. Although varying the SNR across batches is a common practice \cite{nachmani2018}, it is not strictly necessary for effective training \cite{gruber2017}.

\subsection{Pros and cons of this approach}

Training with on-demand data is fast, as modern CPUs and GPUs excel at generating random noisy codewords on the fly. Additionally, it removes the need for storing large datasets, potentially saving tens or even hundreds of gigabytes of storage space. More importantly, the risk of overfitting is significantly reduced since the model is continuously exposed to new, unique examples throughout the training process.

Training with on-demand data offers several advantages but also has a significant drawback: it requires considerably more data compared to using a fixed dataset, where each sample is reused in every epoch. Deep learning practice has shown that stochastic gradient descent converges faster when it repeatedly iterates over the same data. The second, and perhaps more subtle, issue with on-demand data generation is related to the choice of target variables. Decoder models trained with on-demand data are usually trained for \emph{zero-error decoding}, meaning the true error pattern, $\mathbf{e}_\text{chan} = \mathbf{z} - \mathbf{c}$, or, equivalently, the transmitted codeword $\mathbf{c}$, is used as the target during training. However, since any channel decoder, including MLD, inevitably makes errors, training with true error patterns as targets creates a mismatch between the intended function of the model and the way it is trained. This mismatch can ultimately hinder the model's performance. On the other hand, using true error patterns during training avoids the need to include the reference decoder in the  on-demand data generation loop, which would significantly slow down the training process. By using a fixed dataset instead, it is possible to achieve the best of both worlds.

\section{Improving Training With The Help Of Well-Designed Datasets}
\label{sec:datasets}

In this section, we argue that SBND models can be trained to achieve superior performance with fewer training samples by combining several straightforward yet effective strategies, including: 1) utilizing a fixed dataset, 2) selecting alternative target variables, and 3) adopting a distribution of training data different from that induced by the channel model.

\subsection{The benefits of training with a fixed dataset}

In deep learning, the conventional approach involves using a fixed dataset for model training. This approach provides complete control over the training data. In the context of channel decoding, it translates to generating the training data once through offline processing, enabling careful selection of training examples. Additionally, making these datasets publicly available allows other researchers to benchmark their models under identical conditions\footnote{Some of the datasets created for this study are available on the AI4CODE project homepage \href{https://ai4code.projects.labsticc.fr/software}{https://ai4code.projects.labsticc.fr/}. Code to reproduce all results is available at \href{https://github.com/lebidan/sbnd}{https://github.com/lebidan/sbnd}.}. As shown in subsection \ref{subsec:performance_limits}, fixed datasets are also instrumental in analyzing how model performance scales with the training set size for a given learning architecture and model capacity. This leads to a critical question: how should the pairs $(\mathbf{y}, \mathbf{e})$ that constitute the dataset be selected?

\subsection{Training to correct MLD error patterns}
\label{subsec:ml_targets}

To closely approximate MLD, we argue that neural decoders should be trained to replicate the output of an MLD decoder. Thus, we propose replacing the true error patterns, $\mathbf{e}_\text{chan} = \mathbf{z} - \mathbf{c}$, used as targets when training with on-demand data, with the most likely error patterns, $\mathbf{e}_\text{ML} = \mathbf{z} - \mathbf{c}_\text{ML}$, where $\mathbf{c}_\text{ML}$ represents the MLD decision for each received word $\mathbf{y}$. Using a fixed dataset facilitates this replacement without compromising training efficiency. An Ordered-Statistics Decoder (OSD)~\cite{fossorier1995} with a maximum reprocessing order of $i_\text{max} = \lfloor d_\text{min}/4 \rfloor$ was used to obtain $\mathbf{c}_\text{ML}$ and create the datasets used throughout this study. The most straightforward method for constructing such a dataset involves performing a Monte Carlo simulation of the OSD decoder and collecting the resulting error patterns $\mathbf{e}_\text{ML}$. However, as will be discussed in the following subsections, this may not be the most efficient approach.

\subsection{Optimizing the distribution of training data}
\label{subsec:training_distrib}

Monte Carlo methods are known to be inefficient for rare-event simulations. We argue that on-demand generation of training data inherits this same limitation: not all samples are equally beneficial for training neural decoders, as previously highlighted in \cite{beery2020}. Specifically, it is well known that the AWGN channel induces a binomial distribution
\begin{equation}
p_\text{chan}(w) = \binom{n}{w} p_b^w (1-p_b)^{n-w} \quad, \quad w=0,\ldots,n 
\end{equation}
on the Hamming weight $w_H(\mathbf{e}_\text{chan})$ of the true error patterns at the channel output, where $p_b = \frac{1}{2} \text{erfc}(\sqrt{R E_b/N_0})$. Consequently, training with on-demand data or with a dataset constructed from direct Monte Carlo simulation of the BI-AWGN channel results in the model being exposed to a disproportionate number of low-weight error patterns, which are relatively easy to learn, compared to higher-weight error patterns, which are considerably more challenging to assimilate. The key issue is that the model's ability to correct higher-weight error patterns impacts its FER performance, particularly at moderate-to-low SNR. Our objective is to demonstrate that, by carefully selecting the training examples $(\mathbf{y},\mathbf{e})$ and appropriately designing their associated weight distribution $p(w)$ within the dataset, models can achieve improved learning efficiency and better generalization, ultimately resulting in lower test FER \cite{nguyen2024}.

Identifying a provably optimal training distribution for neural decoding appears to be a challenging problem. Instead, we propose comparing various heuristics for constructing the training set. Our objective is twofold: 1) to provide empirical evidence that employing distinct data distributions for training and testing can lead to more data-efficient training and improved model performance, and 2) to identify heuristics that yield favorable results while being simple to implement. We investigate four different methods, all based on Monte Carlo simulation of the OSD decoder. These methods differ in how the samples are selected to form the dataset. The first method is used as a reference, while the other three sampling strategies are designed to focus the learning process on the relevant MLD error patterns that most significantly impact FER performance. A brief overview of each method is provided below. 

\subsubsection{Using the channel noise distribution}
This approach serves as the natural reference against which the other methods will be compared. As described in Subsection \ref{subsec:ml_targets}, the MLD error patterns, $\mathbf{e}_\text{ML}$, are collected directly as they emerge from the decoder output. The MLD decoder effectively transforms the binomial weight distribution $p_\text{chan}(w)$ at its input into a different, more concentrated distribution $p_\text{ML}(w)$, which tends to emphasize lower-weight error patterns.

\subsubsection{Using a uniform distribution of the weights}
We begin by discarding all error patterns $\mathbf{e}_\text{ML}$ with a Hamming weight greater than a specified maximum threshold, $w_\text{max}$, at the MLD decoder output. From the remaining error patterns, we then sample to achieve a uniform weight distribution, denoted as $p_\text{uni-w}(w) = 1/w_\text{max}$ for $w = 1, \ldots, w_\text{max}$, within the dataset.

\subsubsection{Using a biased input noise distribution}
The training examples $\mathbf{y}$ provided to the MLD decoder are generated according to an input distribution $p_\text{chan}^{\text{(is)}}(w)$, which differs from the binomial distribution $p_\text{chan}(w)$ induced by the AWGN channel. Specifically, the noise importance sampling distribution $p_\text{chan}^{\text{(is)}}(w)$ is designed for efficient simulation of MLD decoder performance by encouraging the occurrence of error events near the decision boundaries of the MLD decoder. This distribution is determined following the procedure outlined in \cite{pan2022}. The corresponding weight distribution of error patterns at the MLD decoder output will be referred to as $p_\text{is}(w)$.

\subsubsection{Using a uniform distribution of the syndromes}
Given that the syndrome is integral to the MLD decision rule and also part of the model input, it is reasonable to ensure that all possible (non-zero) syndromes are adequately represented in the training set. A straightforward approach to achieve this is to filter the ML error patterns collected from the output of the OSD decoder, ensuring that all syndrome values are present in the dataset, with an equal number of error patterns for each syndrome. The resulting weight distribution of error patterns will be denoted by $p_\text{uni-s}(w)$.

\section{Numerical Experiments}
\label{sec:experiments}

This section presents numerical results that support the training methodology proposed in Section \ref{sec:datasets}. Most experiments were conducted using the $(31, 21, 5)$ BCH code. The application to longer codes is explored in Subsection \ref{subsec:longer_codes}.

\subsection{Should one train with on-demand data or with datasets?}

\begin{figure}[t!]
\centerline{\includegraphics[scale=0.5]{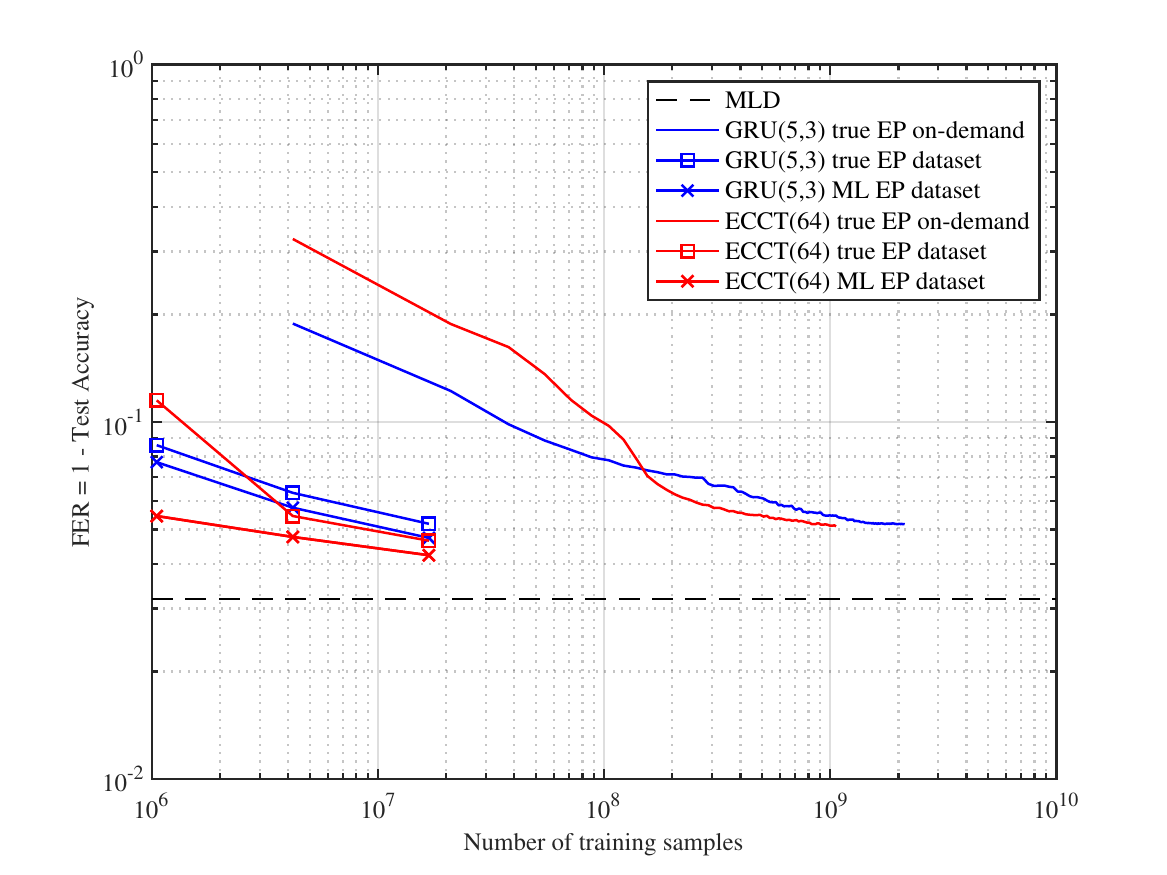}}
\caption{Frame error rate as a function of the number of training samples for different SBND models and training strategies on the $(31,21,5)$ BCH code, at $E_b/N_0=3$ dB (legend: EP=Error Pattern, ML=Maximum-Likelihood).}
\label{fig:sample_complexity_31_21}
\end{figure}

Fig.~\ref{fig:sample_complexity_31_21} illustrates the evolution of the test FER as a function of sample complexity (i.e., the number of training examples) for two SBND decoders trained on the $(31, 21)$ BCH code at $E_b/N_0 = 3$ dB: the GRU$(5,3)$ model and the ECCT$(64,6)$ model. The details of the training setup are provided in the Appendix. This figure compares training with on-demand data versus training with fixed datasets. For the latter, we also compare learning to correct the true error patterns $\mathbf{e}_\text{chan}$ against learning to correct MLD error patterns $\mathbf{e}_\text{ML}$. First, consider the scenario where the model is trained to correct the true error patterns, which is the conventional paradigm. The results in Fig.~\ref{fig:sample_complexity_31_21} show that, for both models, achieving a comparable target FER requires significantly fewer samples—over two orders of magnitude less—when training with a fixed dataset compared to on-demand data. This observation is further supported by the FER vs SNR plots in Fig.~\ref{fig:fer_31_21_true_ml}.\footnote{It is noteworthy that, despite being trained at $E_b/N_0 = 3$ dB, both models generalize well across all other simulated SNR values.}. Specifically, GRU and ECCT models trained on datasets containing 4 million true error patterns perform on par with those trained using 1 billion true error patterns generated on demand. This highlights that training with fixed datasets is substantially more data-efficient than training with on-demand data.

\subsection{Should one learn to correct MLD or true error patterns?}

\begin{figure}[t]
\centerline{\includegraphics[scale=0.5]{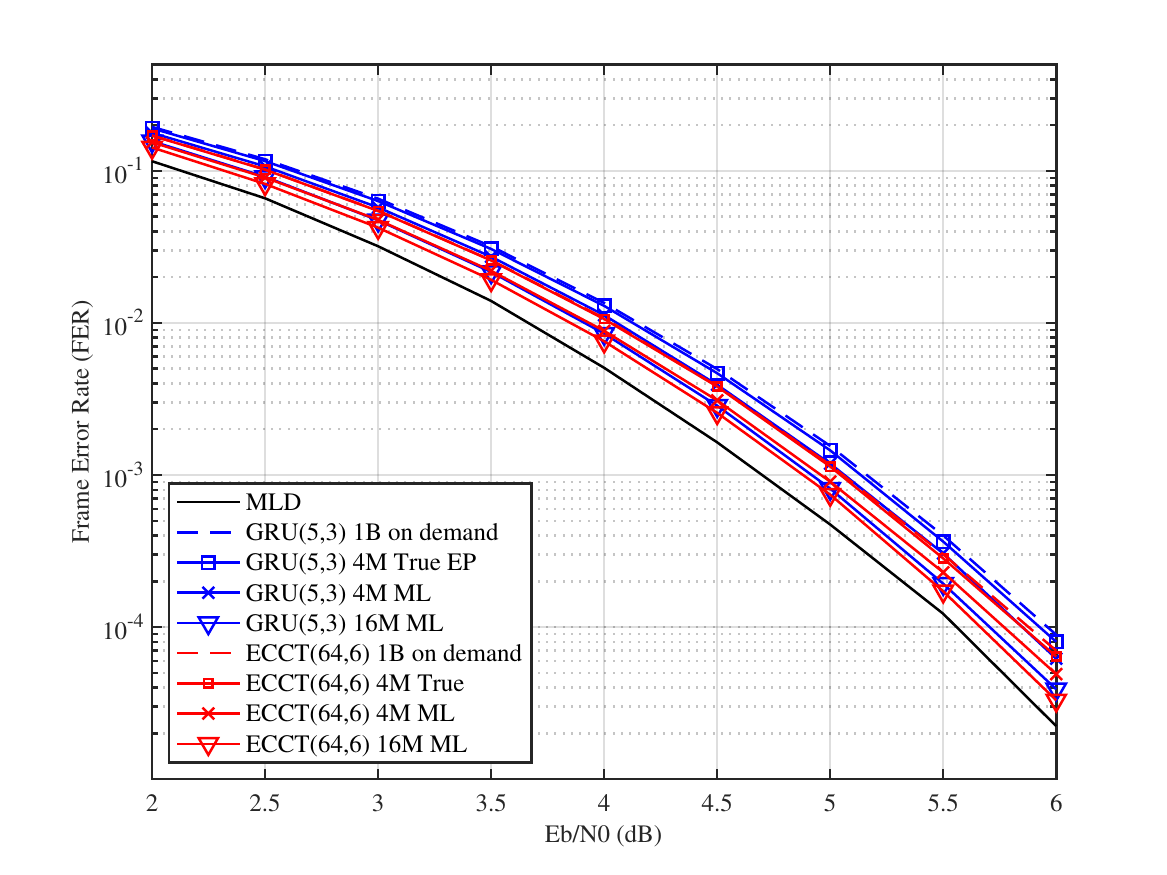}}
\caption{Frame error rate as a function of $E_b/N_0$ for different SBND models and training strategies on the $(31,21,5)$ BCH code.}
\label{fig:fer_31_21_true_ml}
\end{figure}

The results in Fig.~\ref{fig:sample_complexity_31_21} indicate that training models to correct MLD error patterns, $\mathbf{e}_\text{ML}$, consistently yields better performance—reflected by a lower test FER—compared to training on true error patterns, $\mathbf{e}_\text{chan}$. This performance advantage is even more pronounced when using smaller datasets or at lower SNR values. As seen in Fig.~\ref{fig:fer_31_21_true_ml}, the combined advantages of training with a dataset containing 16 million MLD error patterns result in a gain of approximately $0.3$ dB at a FER of $10^{-4}$, compared to the performance of a similar model trained with 1 billion true error patterns generated on demand.

\subsection{Performance limit of architectures and models}
\label{subsec:performance_limits}

Fig.~\ref{fig:sample_complexity_31_21} shows that, for the $(31,21)$ code, the ECCT$(64,6)$ architecture achieves a superior performance vs sample complexity trade-off compared to the GRU$(5,3)$ architecture. Specifically, the ECCT model makes more efficient use of a given number of training examples, except when the dataset size is very small. This observation is further confirmed by the FER plots in Fig.~\ref{fig:fer_31_21_true_ml}. The ECCT model is also considerably smaller, with only 304K parameters compared to 1.7M for the GRU model. Interestingly, both architectures appear to reach a performance ceiling in terms of FER, stabilizing around $0.04$ at the given SNR value, despite increasing sample complexity. This plateau suggests that to closely approach the performance of MLD, a larger model capacity—potentially with more parameters—would be required. Additionally, it is worth noting that for both architectures, a training set size of around 8 to 10 million examples appears to be the threshold beyond which data quality no longer outweighs data quantity. 

\subsection{Impact of the training distribution}

\begin{figure}[t!]
\centerline{\includegraphics[scale=0.5]{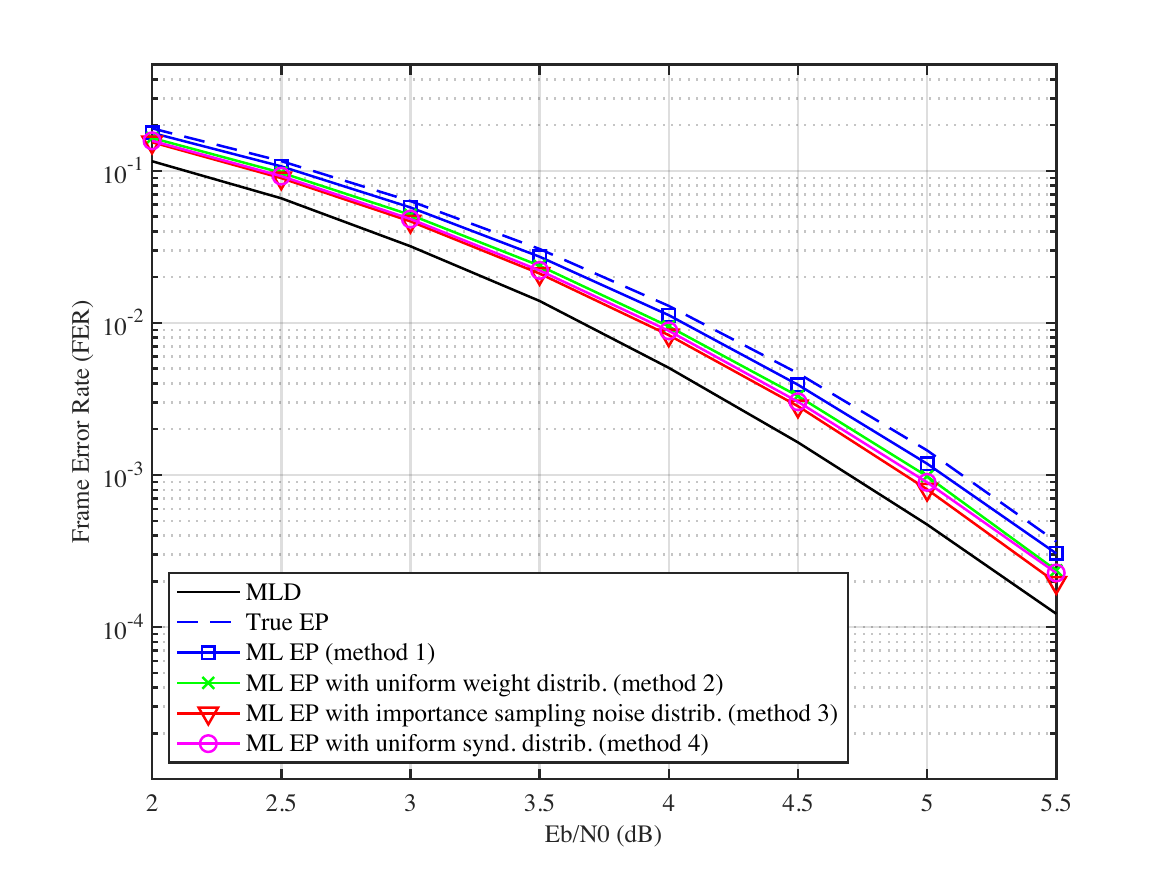}}
\caption{Frame error rate as a function of $E_b/N_0$ for a GRU$(5,3)$ model trained to decode the $(31,21)$ code using different datasets of 4M samples.}
\label{fig:fer_vs_training_distrib}
\end{figure}

\begin{figure}[b!]
\centerline{\includegraphics[scale=0.48]{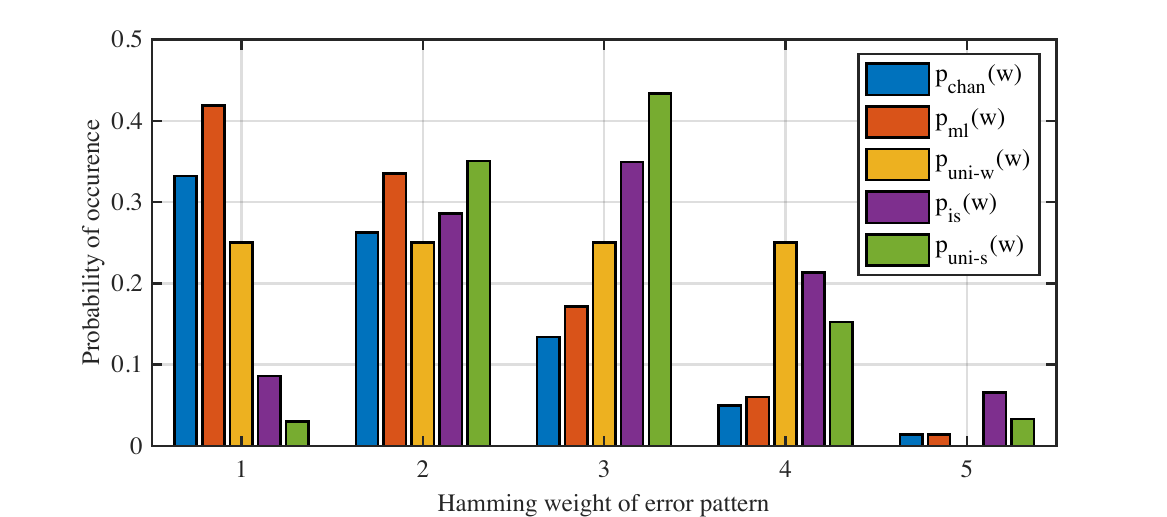}}
\caption{Weight distribution of error patterns obtained with different training set constructions for the $(31,21)$ code at $E_b/N_0=3$ dB.}
\label{fig:wd_training}
\end{figure}

Four different datasets, each containing 4 million MLD error patterns, have been constructed according to the four methods described in Subsection \ref{subsec:training_distrib}, for the $(31, 21)$ BCH code at $E_b/N_0 = 3$ dB. Fig.~\ref{fig:wd_training} presents the weight distribution of error patterns in each dataset, along with the distribution of true error patterns observed at the channel output in the test set. It is evident that methods 3 and 4 place significantly greater emphasis on weight-3 and weight-4 patterns, largely at the expense of weight-1 patterns. Method 2 represents an intermediate between these extremes. A GRU$(5,3)$ model was trained on each of the four datasets, and the results are displayed in Fig.~\ref{fig:fer_vs_training_distrib}. The primary takeaway is that training sets with a distribution differing from the one induced by the BI-AWGN channel, especially methods 3 and 4, yield models with superior performance compared to those trained using true or MLD error patterns collected via standard Monte Carlo simulation. While there is no clearly superior method in this example, the experiment emphasizes the advantages of optimizing the training distribution, especially when working with small datasets. We believe that even more effective dataset construction techniques remain to be discovered.

\subsection{Application to longer codes}
\label{subsec:longer_codes}

\begin{figure}[t]
\centerline{\includegraphics[scale=0.5]{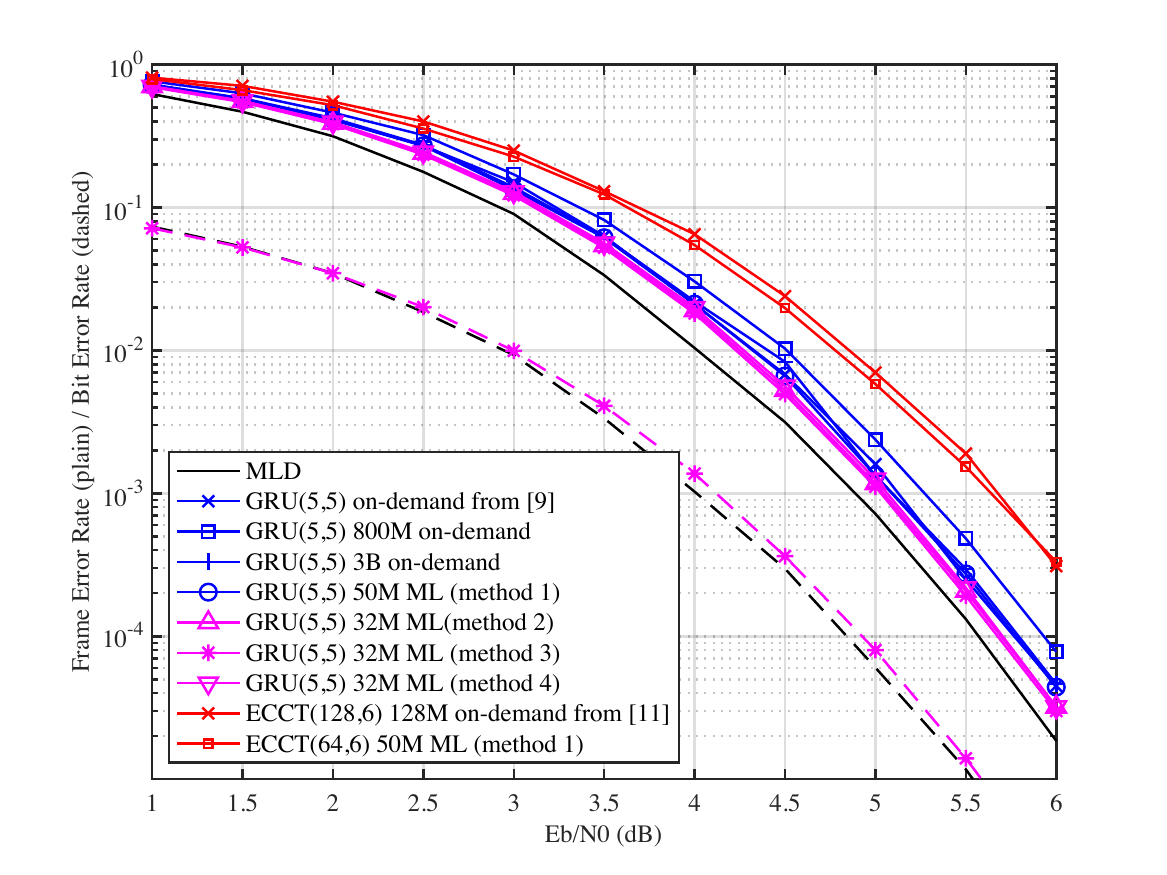}}
\caption{Frame (and bit) error rate as a function of $E_b/N_0$ for different models and training strategies on the $(63,51,5)$ BCH code.}
\label{fig:fer_63_51}
\end{figure}

\begin{figure}[t]
\centerline{\includegraphics[scale=0.5]{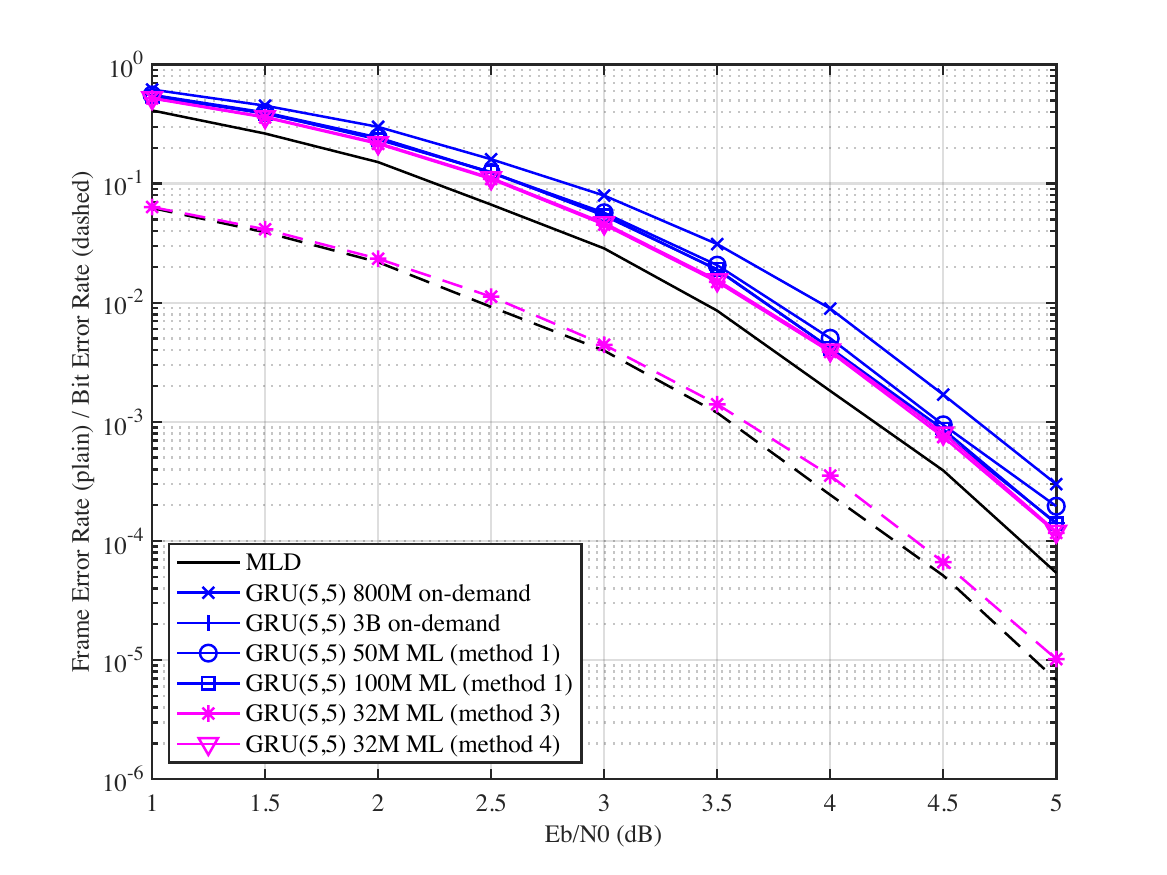}}
\caption{Frame (and bit) error rate as a function of $E_b/N_0$ for different models and training strategies on the $(63,45,7)$ BCH code.}
\label{fig:fer_63_45}
\end{figure}

Figures \ref{fig:fer_63_51} and \ref{fig:fer_63_45} present results for the $(63, 51, 5)$ and $(63, 45, 7)$ BCH codes, respectively. These two codes are frequently used as benchmarks for assessing neural decoders.

First, consider the $(63, 51)$ code. Our GRU$(5,5)$ model trained with 3 billion true error patterns generated on demand achieves performance comparable to that reported in \cite{debonirovella2023}. Notably, the same level of performance can be achieved by training on a dataset consisting of only 50 million MLD error patterns, resulting in a $60\times$ reduction in the number of required training samples. Superior performance can even be obtained using just 32 million samples when employing the optimized training distributions described in Subsection \ref{subsec:training_distrib}. Also shown in the figure is the performance reported in~\cite{choukroun2022} for an ECCT$(128,6)$ model trained with 128 million true error patterns generated on demand. Interestingly, our smaller ECCT$(64,6)$ model, trained with a dataset of 50 million MLD error patterns, slightly outperforms this larger model, thereby reducing both model complexity—from 1.7M to 310K trainable parameters—and sample complexity. However, it is also worth noting that neither of these two ECCT models matches the performance of the GRU$(5,5)$ on this code.

A similar trend is observed with the $(63, 45)$ code. A GRU$(5,5)$ model trained with a fixed dataset of 100 million error patterns achieves performance comparable to that of a model trained on 3 billion true error patterns generated on demand. Furthermore, slightly better performance can be attained using a dataset of only 32 million examples with the sampling strategies from Subsection \ref{subsec:training_distrib}. Increasing the size of the optimized training set to 64 million yields only marginal improvements, not shown here. Similar to what was observed with the $(31,21)$ and $(63, 51)$ codes, we appear to be reaching a performance ceiling for the GRU$(5,5)$ model here as well.

On a final note, we have also shown in Fig.~\ref{fig:fer_63_51} and \ref{fig:fer_63_45} the BER performance of our best models, to demonstrate that, while we have not yet fully bridged the gap to MLD at the codeword level, the BER performance is nearly equivalent.

\section{Conclusion}
\label{sec:conclusion}

This research was motivated by the need to understand why SBND decoders often fall short of achieving MLD performance for many codes of practical interest. Our primary finding indicates that greater emphasis should be placed on the training data. Specifically, we have demonstrated that by using fixed datasets, it is possible to train existing models to achieve comparable or even superior performance levels with significantly fewer training samples compared to the on-demand data generation paradigm commonly used in previous studies. Furthermore, we have shown that training models to correct MLD error patterns rather than true error patterns leads to improved model performance. We have also demonstrated that using a training distribution different from the one naturally induced by the channel can provide substantial benefits in terms of both data efficiency and generalization error. Importantly, most of these findings also extend to model-based decoders, such as neural BP.

\appendix

All models were trained for a maximum of 256 epochs with a batch size of 4096 samples. The AdamW optimizer was used, with a weight decay factor of $0.02$ for the GRU models and $0$ for the ECCT models, in conjunction with the ReduceLROnPlateau schedule. The initial learning rate was set to $0.001$ for the length-31 code and $0.0005$ for the length-63 codes. Dropout proved crucial in preventing overfitting, particularly when training with small fixed datasets. Specifically, we applied a dropout rate of $0.2$ for the GRU models and $0.01$ for the ECCT models; however, no dropout was applied when training with on-demand data. Training data was generated at an $E_b/N_0$ value of $3$ dB for the $(31, 21)$ and $(63, 51)$ codes, and $2$ dB for the $(63, 45)$ code.

\section*{Acknowledgment}

The authors are grateful to G. De Boni Rovella and M. Benammar for their help in reproducing results from \cite{bennatan2018}.

\end{document}